# DISTRIBUTED READABILITY ANALYSIS OF TURKISH ELEMENTARY SCHOOL TEXTBOOKS

**Betul Karakus**
Computer Engineering Department, Firat University 23100 Elazig, Turkey

**Galip Aydın**
Computer Engineering Department, Firat University 23100 Elazig, Turkey

**Ibrahim Rıza Hallac**
Computer Engineering Department, Firat University 23100 Elazig, Turkey

**ABSTRACT**

*The readability assessment deals with estimating the level of difficulty in reading texts. Many readability tests, which do not indicate execution efficiency, have been applied on specific texts to measure the reading grade level in science textbooks. In this paper, we analyze the content covered in elementary school Turkish textbooks by employing a distributed parallel processing framework based on popular MapReduce paradigm. We outline the architecture of a distributed Big Data processing system which uses Hadoop for full-text readability analysis. The readability scores of the textbooks and system performance measurements are also given in the paper.*

***KEYWORDS***— *Readability, Hadoop, Textbook*

## I. INTRODUCTION

The difficulty (or ease) of a selected text can be expressed as its readability; which in turn can be used to classify reading materials into different grade levels. Instructors can make use of readability scores to find suitable teaching materialsAside from education, readability applications have some potential usage areas such as business publications, complex financial reports, online media [3]; health care [4], military agencies'enlistment applications and technical manuals and web pages [5].

Various readability formulas have been proposed to measure the level of text difficulty. The FleschReading Ease [6], Flesch-Kincaid Grade Level [7], SMOG Index [8], Gunning Fog Index[9], Automated Readability Index [10] and Dale-Chall readability formula [11] are among the best known readability formulas. These popular formulas are widely used to improve textbooks, health literature, business and finance publications, military and governmental documents, web contents and, so forth.The readability formulas, developed by various researchers since the 1920s, aim to assess the difficulty of a given text.

Before these studies, the difficulty of the text was tried to be determined with a method based on reading comprehension. However, Fletcher [12] has emphasized the fact that the assessment of reading comprehension is difficult because it is not a process that can be directly observed.

Readability assessment plays an important role in document analysis. Web pages on the internet contain large amount of valuable information.The number of online documents has been increasing with incredible growth rates. However, the users do not usually have the means to find suitable reading materials according to their reading grade level. On the other hand, analyzing large number of documents require high computational power and storage space. English Wikipedia which has 5 TB of data as set of documents is one of the best examples.Similarly, textbooks or their electronic copies require powerful computers for efficient analysis. Therefore a big data solution is required to assess the full-text readability of textbooks.

Big data has many challenges on several aspects like variety, volume, velocity and veracity. Variety refers to unstructured data in different forms, velocity refers to how fast the data is generated and how fast they need to be analyzed and veracity refers to the trustworthiness of data to be reliable for crucial decisions [13]. There are several distributed solutions to handle big data. The most well-known is MapReduce framework, which was published by Google [14].Most of the research on readability analysis have focused on runtime performance of their implementations. In this study, we present our Distributed Readability Analysis System, which is based onHadoop Distributed Computing Framework for analyzing the readability of Turkish elementary school textbooks used by students from grade 5 to 8.

The rest of the paper is structured as follows. Section II presents the literature survey, Section III gives a review of readability analysis as needed for following discussion. Section IV presents our Distributed Readability Analysis System. Section V presents the performance evaluation. Finally, Section VI gives conclusions and explains future works.

## II. RELATED WORK

Research on text readabilitybegan in the last quarter of 19th centurywith "Analytics of Literature"by L.A. Sherman [15]. He drew attention to the importance of average sentence length and demonstrated that shorter sentences increase readability. Readability studies proposed in the following century primarily focused on the development of readability formulas to select appropriate textbooks according to ability of the students. Lively and Pressey [16] developed first readability formula to measure and reduce the vocabulary burden of textbooks. Flesch published his popular Reading Ease formulato measure reading materials after a number of studies on English readability and other popular formulas were published from the 1940s up to the middle of 1990s (see Table 1 for details).

More recently, readability formulashave been usedfor measuring the reading grade level of the textbooks and the validity of the formulas have been evaluated not only in Englishbut also in many different languages. However,it has been observed that classic readability formulas such as Flesch Reading Ease are only effective for English languages [17].On the other hand, most of the languages are similar to each other in respect to general lexical features such as average sentence length, average number of syllables, average number of words per sentence and average number of hard words (more than three syllables). Kuo et al. investigated the problem of readability analysis in Taiwanese texts and classified short essays for high school students using linguistic features [18]. The analysis results of [18] show that Flesch-Kincaid Grade Level, SMOG and Automated Readability Index scores could not achieve good results for predicting readability.

Another study on Polish language [19] has implemented and evaluated Gunning Fog Index and Flesch-based Pisarek method using specific lexical items in Polish texts. In order to assess readability of Thai text for primary school students, Daowadung and Chan [20] has developed a technique to predict readability of Thai textbooks using word segmentation and TF-IDF calculations. Similarly, TF-IDF vectors has been constructed to determine readability of primary school Chinese textbooksin [21]. The experimental results showed that the proposed method is effective for lower grades, but not effective for middle grades.

François and Fairon [22] have implementedtheir French readability formula and classification model relative to the set of textual features includingthe lexical, syntactic, and semantic features on a textbook corpus. Aluisio et al [23] have alsostudied the readability assessment for Portuguese texts using the linguistic structure of the texts. Other research for Japanese text readability has proposed a readability measurement method based on textbook corpus, which consists of 1.478 sample passages extracted from 127 textbooks [24]. They use a simple Perl program to analyze the textbook readability. The Lasbarhetsindex Swedish Readability Formula(LIX) is also among popular readability formulas to measure the difficulty of reading a foreign text [25]. While other popular formulas count the number of syllables for average word length, LIX uses the number of letters as shallow feature, which is the traditional feature used for readability analysis.Sjöholm [26] has developed a java module taking a Swedish text as input to evaluate and measure LIX formula.

Although classic readability formulas are widely used on texts written in English, semantic languages such as Arabic need to generate new readability formulas for selecting appropriate textbooks [27]. Al-Khalifa and Al-Ajlan [28] has developed a java program to calculate a vector of values for the four features: average sentence length, average word length, average syllables per word and word frequencies. Their corpus collected from Saudi Arabia schools consists of 150 texts with 57089 words.

Early work on Turkish readability began in 1990s. Two popular formulas have been proposed to measure Turkish text readability: Atesman [29] has proposed first readability formula, which is the adaptation of Flesch Reading Ease Formula (see Table 1 for details).Cetinkaya [30] has presented second readability formula with three readability levels: highlevel (10th, 11th, 12th grade), intermediate reading (8th and 9th grade) and elementary reading (5th, 6th and 7th grade).

**TABLE I Readability approaches in literature**

| Readability Formula | Index | Formula Content | Shallow Features |
|---|---|---|---|
| Flesch Reading Ease | 6 | $206.835 - 1.015\left(\frac{total\ words}{total\ sentences}\right) - 84.6\left(\frac{total\ syllables}{total\ words}\right)$ | Average Sentence Length, Syllables-based |
| Flech-Kincaid Grade Level | 7 | $0.39\left(\frac{total\ words}{total\ sentences}\right) + 11.8\left(\frac{total\ syllables}{total\ words}\right) - 15.59$ | Average Sentence Length, Syllables-based |
| SMOG | 8 | $1.0430\sqrt{hard\ words \times \frac{30}{sentences}} + 3.1291$ | Hard words (more than three syllables) |
| Gunning Fog | 9 | $0.4 \times \left[\frac{words}{sentences} + \left(100 \times \frac{hard\ words}{words}\right)\right]$ | Average Sentence Length, Hard words |
| Automated Readability | 10 | $4.71\left(\frac{letters}{words}\right) + 0.5\left(\frac{words}{sentences}\right) - 21.43$ | Average Sentence Length, Character-based |
| Dale-Chall | 11 | $0.1579\left(\frac{difficult\ words\ (not\ in\ list)}{words} \times 100\right) + 0.0496\left(\frac{words}{sentences}\right)$ | Average Sentence Length, Wordlist-bases |
| Atesman [20] | 27 | $198.825 - 40.175\left(\frac{total\ syllables}{total\ words}\right) - 2.610\left(\frac{total\ words}{total\ sentences}\right)$ | Average Sentence Length, Syllables-based |

Okur and Arı [31], have analyzed and compared 298 selected texts in Turkish textbooks for grades 6-8 using Atesman and Cetinkayaformulas. In their study, selected texts were divided into two categories, namely informative reading texts and narrative reading texts. It is observed that the readability level of the informative reading texts is more difficult than of the narrative ones because the average sentence length and word length in narrative reading texts are shorter when compared with the informative reading texts. Guven [32] has also studied on readability of texts in Turkish textbooks used by foreigners.

All of these readability studies point to the fact that most of the readability studies have been done on sample passages or selected units of textbooks rather than the whole of the textbooks. The studies using text samples need to carefully select the sample text in order to avoid calculating incorrect readability scores. That is, the accuracy of readability scores depends on the selected sample texts. Furthermore, due to lack of a readability system to analyze whole textbooksin a short time and due to lack of Turkish textbook corpus, the readability implementations are not used by the educational institutions in Turkey.

### III. READABILITY ANALYSIS

The variables used in the readability formulas show us the skeleton of a text [33]. Traditional readability formulas have been built on many variables, which affect the reading difficulty. These variables include the following landmark features:

- Readability is defined as "the ease of reading words and sentences" by Hargis [34]. That is, the readability of text depends on the clarity of words and sentences. Themost common features to predict word and sentence complexitiesare average sentence length, average word length, number of easy-hard words and number of simple sentences.

- The vocabulary diversity or the number of different words in a text, plays an important factor on reading difficulty. Dale-Chall(see Table 1 for formula) compiled a list of 3000 easy words by comparing the number of different words.

- Readability challenges include not only lexical difficulty, but also structural difficulty. Structural features used in various readability formulas are the number of propositions, kinds of sentences and prepositional phrases. However, these formulas lead Kintsch and Miller [35] to the conclusion that lexical features such as word and sentence length provide stronger measurement for text difficulty, when compared to structural factors related to mental properties of a reading text.

Developmentsin computer software accelerated readability analyses and today, readability studies are more popular thanin the past. Microsoft Word, the most popular word processing software, can automatically calculate the readability of text using the Flesch Reading Ease and the Flesch–Kincaid Grade Level formulas. Readability analyses have undergone a major change in the last twenty years.The ATOS Readability formula [36] isbased on three key features including average sentence length, average word length and average word difficulty level, has been one of the most important changes. While first graded vocabulary list was consist of almost 24.000 words in 2000, the list underwent a major update adding 75.000 new words in 2013. The study also emphasizes that the readability analyses in the past lead to incorrect results due to the use ofonly small samples of text rather than whole books.

As the amount of web pages and electronic copies of textbooks on the Internetgrow rapidly, it becomes difficult to analyze these documents using traditional applications for finding suitable materials for individual readers and students at different grade levels[37]. Hadoop-based electronic book conversion system proposed by [38] presents a distributed solution to process large numbers of electronic books. The study indicates that it takes very long processing times, when the proposed system is performed on a personal computer.

The main requirements in analyzing large documents are high computing power and storage capacity. Big data technologies support distributed storage and data processing, which allows us to use commodity hardware for parallel, and distributed computing. This study builds a distributed framework to analyze textbook readability using big data technologies.

## IV. DISTRIBUTED READABILITY ANALYSIS SYSTEM

In this paper, we propose a Distributed Readability Analysis System (DRAS) for Turkish elementary school textbooks using Hadoop framework.The overview of the proposed system is given in Fig. 1. The system architecture is divided into three parts namely data conversion, data pre-processing and data analysis. The main goal of the proposed system is to provide an efficient readability analysis system to help educators and parents in finding appropriate reading materials for elementary school students. We use two keyfeatures,average sentence length and average word length, which are used by Atesman Readability formula. Furthermore, we calculate the total number of distinct words in each textbook for testing the accuracyof the readability formulas.

### A. Textbook Data Conversion

The official website of The Republic of Turkey's Ministry of National Education (MoNE) presents Turkish textbooks as PDF files. We first convert the PDF textbooks into text files using Apache Tika [39], which provides textual analysis software that runs on top of the Hadoop framework. MapReduce jobs were run to transform PDF files into text files and they were stored locally. Religious culture textbooks were not successfully converted due to the use of Arabic texts and 12 PDF files which only contained images were not converted. The execution time efficiency of the data conversion is detailed in Section V as Table IV.

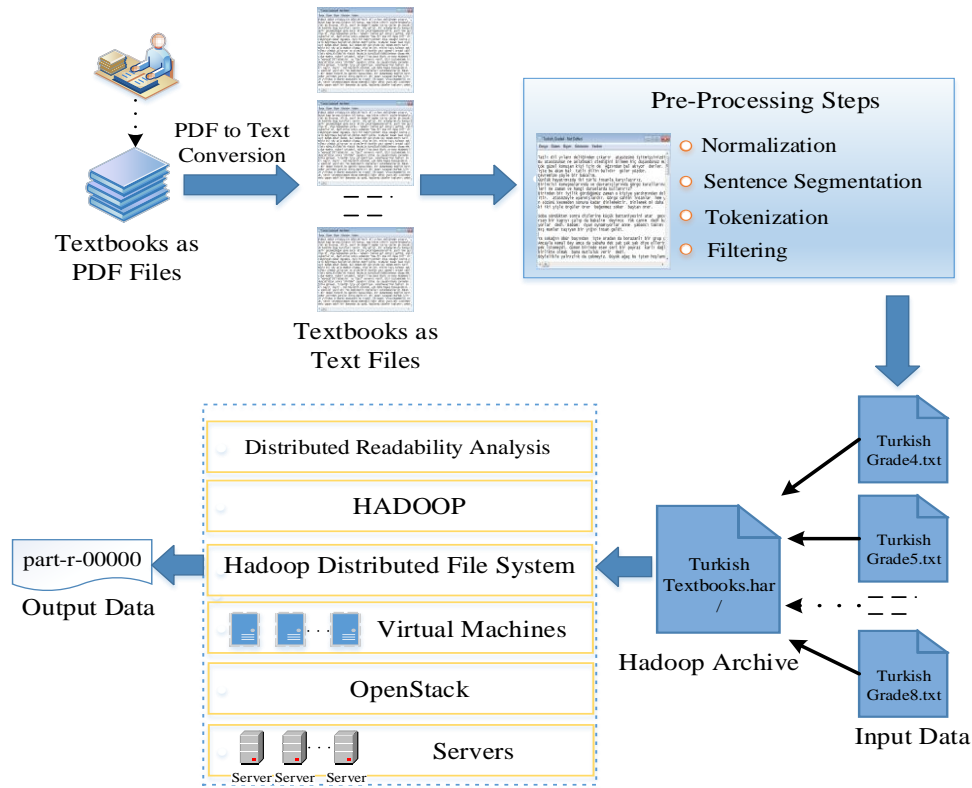

**Fig.2 An overview of Distributed Readability Analysis System**

*B. Textbook Data Pre-processing*

Before analyzing the Turkish textbooks, we have performed the following pre-processing steps.

- Normalization, which is the conversion of character set to UTF-8 Unicode encoding.

- Sentence segmentation, which is the segmentation of document content into sentences. We have used Zemberek, Turkish NLP Library [40] for this step.

- Tokenization, which is the process of breaking up a text into tokens. In this step, multiple whitespace characters are replaced by a single whitespace character.

- Filtering, which is the removal of special characters (+,-,*, #, /, \, $, =, &), numbers, stop words (i.e. words in meaningless words category such as that, this, it, etc.)and author names.

Most of the Turkish readability studies in literature pay no attention to the impact of stop words in their readability assessments. We eliminate the stop words in the selected textbooks to overcome this major shortcoming. Furthermore, we construct a set of Turkish stop word by determining term frequencies of each word. The general approach for creating a stop word list is to sort the words according to their term frequency, which is the number of times the word appears in the document. Stop words are high frequency terms, that is, they are the most common words, which are less significant than other words. Therefore, we remove the stop words from the texts.

*C. Textbook Data Analysis*

We use Hadoopto analyze the input data, which is the preprocessed textbook stored in HadoopDistributed File System (HDFS). Hadoop [41] is the most widely used system in big data analysis. Traditional readability methods performed on a single machine typically usessmall and controlled datasets such as sample passages or units of textbooks. However full textbook data is larger and noisy to be processed using small computational hardware. In order to avoid this problem, we useHadoop platform to analyze the readability of Elementary school textbooks.

The Hadoop platformprimarily provides two main abilities: distributed computing and parallel processing. For these, Hadoop Distributed File System (HDFS) and MapReduce jobs are used. The data are stored in HDFS and MapReduceuses these data as input. MapReduce jobs controlled by a master node are splinted into two functions as Map and Reduce. The Map function divides the data into a group of key-value pairs and the output of each map tasks are sorted by their key. The Reduce function merges the values into output data, which is stored in HDFS.In our Distributed Readability Analysis System, TheMapReduce phases are describedas follows:

1) *Mapper Phase*: The mapper parses each block of input data, which are the textbook contents from HDFS. The first mapper gets the number of sentences, words and syllables (shown in Table II) and then writes the following key value pair: <file name, textbook content>. The second mapper parses each wordand counts how many times it occurs in a textbook file. Then the mapper emits the numeric value "1" for each word and it writes to the reducer the following key value pairs: < (word, filename), 1>. The keys includethe distinct words (shown in Table II) in the textbook files and the values include a list of emitted numeric values for each word. In the mapper phase, the words including special characters, digits and the stop words are removed to filter the textbook data and prevent any occurrence of them inthe calculation of the distinct words.

2) *Reducer Phase:*The reducer gets the results from the mapper as input data and sums up the number of sentences, words and syllables. Then, it divides the number of sentences into number of words to measure average word length and also divides the number of words into number of syllables to measure average sentence length. The next step, the reducer calculates Atesman readability scores and writes the following key value pairs to the output :|<file name, list of (textbook content)>. The second reducer sums upthe number of occurrence of the distinctwords in each textbook file and it writes to the output data using <filename> as the key, <list of (n)> as the value, which includes list of the total number of distinct words in each textbook data.

After implementation of the proposed system, the graphical results of average sentence length, average word length and readability scores are shown in Fig. 2, Fig. 3 and Fig. 4, respectively. As the grade level of the textbooks increase, average sentence and word length has also increased in proportion to the readability level. After all, we have achieved high accuracy in predicting readability of Turkish elementary school textbooks. However the readability of $4^{th}$- $5^{th}$ grades and $6^{th}$ -$7^{th}$ grades represents the change betweenscores 60 and 80, whereas the readability score of the $8^{th}$grade has surprisingly reached score 40.

**TABLE II FEATURE DISTRIBUTION OF TURKISH TEXTBOOKS**

| Textbooks | Word Count | Sentence Count | Syllable Count | Distinct Words |
|---|---|---|---|---|
| $4^{th}$ Grade | 14917 | 1994 | 38396 | 4243 |
| $5^{th}$ Grade | 17484 | 2309 | 45293 | 5095 |
| $6^{th}$ Grade | 18232 | 2315 | 47726 | 5195 |
| $7^{th}$ Grade | 20517 | 2482 | 52692 | 5835 |
| $8^{th}$ Grade | 16895 | 1694 | 54680 | 5798 |

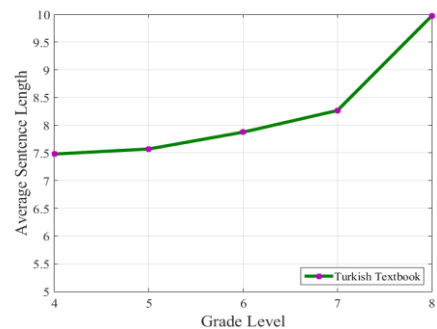

**Fig.3 Average number of syllables per word**

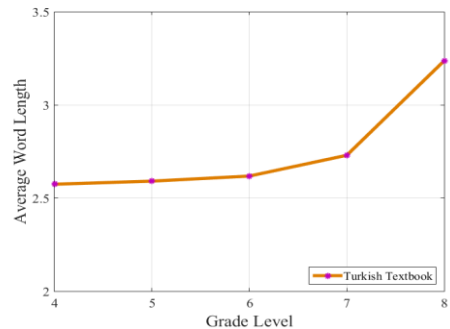

**Fig.4 Average number of words per sentence**

## V. PERFORMANCE EVALUATION

In order to analyze the readability of primary school Turkish textbooks described in this paper, we have created to different applications. The first one is a Java console application for testing the performance of a single machine traditional programming approach. The console application was performed on aPC with four-core 4.1 GHz Intel processor and 16 GB of main memory and used Ubuntu 14.04 LTS as the operating system.

The second application used the distributed Hadoop based system. We used OpenStack for creating a 10-node Hadoop cluster. Performance comparison for the two application is given in Table III. The tests, which running on Hadoop cluster, demonstrated that the execution time of the MapReduce application decreases as the number of the textbook increases and offers better performance than the console application, if the number of textbook is over 2.

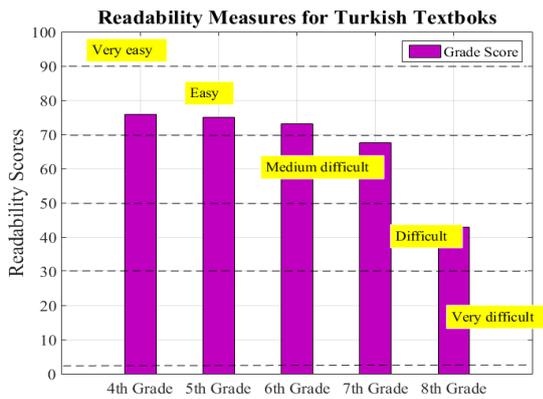

Fig. 5. Predicting readability of elementary school textbooks

**TABLE III PERFORMANCE COMPARISON**

| Number of Textbook | Console Application (sec) | MapReduce Application(sec) |
|---|---|---|
| 1 | 1.5 | 2 |
| 2 | 6.5 | 3 |
| 5 | 141.692 | 5 |
| 10 | Out of memory | 10 |
| 100 | Out of memory | 57 |

**TABLE IV TEXTBOOK DATA CONVERSION ON HADOOP PLATFORM**

| Number of Textbook | File Size (GB) | Running Time | Total Time (min) |
|---|---|---|---|
| 10 | 1.3 | 15:21:24-15:25:39 | 4.25 |
| 20 | 2.5 | 15:33:09-15:39:25 | 6.26 |
| 50 | 5.6 | 15:45:32-15:54:19 | 8.78 |
| 100 | 11.2 | 16:00:21-16:14:35 | 14.23 |

## VI. CONCLUSIONS

This study focus on how the readability system can provide high performance distributed execution for the elementary school textbook based key features such as average sentence length, average word length and distribution of distinct words. For this purpose, the proposed system is evaluated and tested on HadoopMapReduce platform as the distributed execution engine. After having built the application platform, we compared the performance of our distributed readability analysis. The performance results show that our distributed readability analysis system performed on a large number of Turkish textbooks is feasible and efficient.

The main challenge to our study is the lack of the Turkish textbook corpus. We dynamically constructed a distributed readability analysis system on the top level of Hadoop cluster because we intend to expand our textbook corpus in the feature study. Other future work will focus on improving of Turkish readability measures.